\documentclass[%
 reprint,
superscriptaddress,
 amsmath,amssymb,
 aps,
]{revtex4-2}

\usepackage{graphicx}% Include figure files
\usepackage{dcolumn}% Align table columns on decimal point
\usepackage{bm}% bold math

\usepackage[dvipsnames]{xcolor}
\usepackage{stackrel}
\usepackage{sidecap} 

\usepackage{amsfonts}
\usepackage{physics}

\usepackage[normalem]{ulem}

\renewcommand{\arraystretch}{1.5}
\usepackage{tabularx}

\usepackage[hidelinks]{hyperref}% add hypertext capabilities

\renewcommand*{\thefootnote}{\fnsymbol{footnote}}

\newcommand{\Dbulk}{D_0}
\newcommand{\Dself}{D_{\rm self}}
\newcommand{\Dcoll}{D_{\rm coll}}

\newcommand{\Lbox}{L}

\newcommand{\DN}{\Delta N}
\newcommand{\DNtt}{\brak{\DN^2(t)}}
\newcommand{\brak}[1]{\langle #1 \rangle}
\newcommand{\Nmean}{\brak{N}}

\newcommand{\Nvar}{\text{Var}(N)}

\newcommand{\SkZ}{S(k\mathbin{=}0)}

\newcommand{\largeL}{L \rightarrow \infty}
\newcommand{\smallL}{L \rightarrow 0}
\newcommand{\smallk}{k \rightarrow 0}
\newcommand{\largek}{k \rightarrow \infty}
\newcommand{\DlargeL}{D(\largeL)}
\newcommand{\DsmallL}{D(\smallL)}
\newcommand{\Dsmallk}{D(\smallk)}

\newcommand{\Dsmallkt}{D(\smallk, t)}
\newcommand{\Dlargekt}{D(\largek, t)}

\usepackage{babel}

\usepackage{xr}
\usepackage[separate-uncertainty=true]{siunitx}

\begin{document}

\preprint{APS/123-QED}

\title{Measuring collective diffusion coefficients}
\title{Revealing collective hydrodynamic interactions via counting particles in boxes}

\title{Measuring collective diffusion to resolve hydrodynamic interactions}

\title{New strategies to measure collective diffusion coefficients}
\title{Measuring collective diffusion properties: \\ Counting and Fourier-based methods}
\title{Measuring collective diffusion coefficients by counting particles in boxes} % I think we can stress that the counting part is the important one and say in the abstract we of course review standard methods. I'm honestly flexible on the title.


\author{Adam Carter}
% \affiliation{Sorbonne Universit\'{e}, Physicochimie des Electrolytes et Nanosyst\`{e}mes Interfaciaux, F-75005 Paris, France}
\affiliation{CNRS, Sorbonne Universit\'{e}, Physicochimie des Electrolytes et Nanosyst\`{e}mes Interfaciaux, F-75005 Paris, France}

\author{Eleanor K. R. Mackay}
\affiliation{Physical and Theoretical Chemistry Laboratory, South Parks Rd, Oxford, OX1 3QZ UK}

\author{Brennan Sprinkle}
%\email{bsprinkl@mines.edu}
\affiliation{Applied Math and Statistics, Colorado School of Mines, 1500 Illinois St, Golden, CO 80401}

\author{Alice  L. Thorneywork}
%\email{alice.thorneywork@chem.ox.ac.uk}
\affiliation{Physical and Theoretical Chemistry Laboratory, South Parks Rd, Oxford, OX1 3QZ UK}
%\affiliation{Cavendish Laboratory, Department of Physics, University of Cambridge, JJ Thomson Avenue, Cambridge CB3 0HE, UK}

\author{Sophie Marbach}
\email{sophie.marbach@cnrs.fr}
\affiliation{CNRS, Sorbonne Universit\'{e}, Physicochimie des Electrolytes et Nanosyst\`{e}mes Interfaciaux, F-75005 Paris, France}

\date{\today}

\begin{abstract}
% Resolving collective diffusion properties, \textit{i.e.} how fast density inhomogeneities relax, is a widespread experimental and numerical challenge, as it requires resolving motion over large space and time scales. Investigations often rely on probing the relaxation of density correlations at equilibrium in Fourier space, fraught with difficulties pertaining to accurately taking Fourier transforms. Out-of-equilibrium investigations are especially hard to implement experimentally.  
% Here, we introduce a novel approach to resolve collective relaxation of particles, simply by counting the number of particles $N(t)$ in virtual observation boxes of an experimental image, or a simulation dataset, at equilibrium -- the so-called ``Countoscope''. We investigate, in a model 2D colloidal suspension, the decorrelation timescale of the correlation of particle numbers $\langle N(t) N(0) \rangle$. We find that larger box sizes are increasingly sensitive to collective effects. The collective diffusion coefficient can be inferred in the limit of large boxes, typically 10 times a particle's diameter. 
% Unlike Fourier transforms which struggle with the finite field of view of an image, counting harvests finite observation windows. We hope our approach will pave the way to rationalize collective properties of more complex suspensions, including unravelling hydrodynamic signatures or the role of specific interactions in active matter.

The collective diffusion coefficient $\Dcoll$ is a key quantity for describing the macroscopic transport properties of soft matter systems. However, measuring $\Dcoll$ is a fundamental experimental and numerical challenge, as it either relies on nonequilibrium techniques that are hard to interpret or, at equilibrium, on Fourier-based approaches which are fraught with difficulties associated with Fourier transforms.
In this work, we investigate the equilibrium diffusive dynamics of a 2D colloidal suspension experimentally and numerically. We use %the recently introduced 
a ``Countoscope'' technique, which analyses the statistics of particle number counts $N(t)$ in virtual observation boxes of a series of microscopy images at equilibrium, to measure $\Dcoll$ for the first time. We validate our results against Fourier-based approaches and establish best practices for measuring $\Dcoll$ using fluctuating counts. %Furthermore, we review the limitations of both counting and Fourier techniques. 
We show that Fourier techniques yield inaccurate long-range collective measurements because of the non-periodic nature of an experimental image, yet counting exploits this property by using finite observation windows.
Finally, we discuss the potential of our method to advance our understanding of collective properties in suspensions, particularly the role of hydrodynamic interactions.
\end{abstract}

\maketitle

%%%%%%%%%%%%%%%%%%%%%%%%%%%%%%%%%%%%%%%%%%%%%%%%%%%%%%%%%%%%%%%%%%%%%%%%%%%%%%%%%%%%%%%%%%%%%%%%%%%%%%%%%%%%%%%%
%\section{\label{sec:intro} Introduction}
%%%%%%%%%%%%%%%%%%%%%%%%%%%%%%%%%%%%%%%%%%%%%%%%%%%%%%%%%%%%%%%%%%%%%%%%%%%%%%%%%%%%%%%%%%%%%%%%%%%%%%%%%%%%%%%%

Understanding the motion of an ensemble of particles, or collective motion, is a fundamental puzzle in soft matter. Outstanding questions in this area range from determining how molecules traverse a porous matrix~\cite{ariskina2022free,obliger2016free,falk2015subcontinuum} to learning how interactions between active or living particles trigger spontaneous group motion~\cite{cates2015motility,liebchen2017phoretic,liebchen2021interactions,dijkstra2021predictive}. 
Here, a canonical example of collective motion is collective diffusion~\cite{dhont1996introduction}. Yet for this most fundamental case we lack a full understanding even for systems with the simplest interactions, in part due to challenges associated with resolving collective dynamic properties from experimental microscopy data. 
%\alice{To pick up a point I make later on - nothing is said now in the introduction that we target doing this for microscopy/from trajectories if you want to make the case also for simulations}

\begin{figure}[h!]
    \centering
    \includegraphics[width=0.99\linewidth]{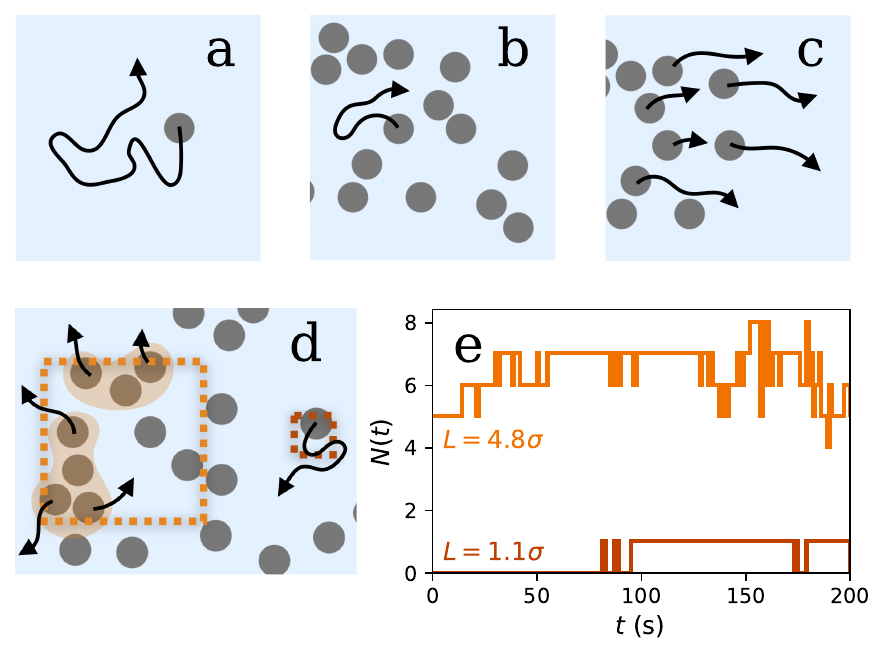}
    \caption{\textbf{Inferring collective diffusion properties from counts}. Diffusion properties can refer to (a) the bulk or free diffusion coefficient of a particle $D_0$ suspended in a fluid, (b) the self-diffusion of a particle $D_{\rm self}$ in a suspension or (c) the collective diffusion coefficient of the suspension $D_{\rm coll}$, that describes the relaxation of a particle density gradient. (d) Here, we show how to measure $D_{\rm coll}$ from the relaxation of groups of particles at equilibrium in large virtual observation boxes on an image (orange). Counts in small boxes (red) probe individual motion. (e) The number of particles $N(t)$ in a box fluctuates due to individuals or groups of particles diffusing in and out of the box. }
    \label{fig:intro}
\end{figure}

To understand the questions and challenges pertaining to the study of collective diffusion, we start by a brief overview of our understanding of self diffusion in particle suspensions. %, and then turn back to collective diffusion. 
Following the seminal works of Stokes and Einstein~\cite{einstein1905molekularkinetischen}, we understand the diffusion of a single particle suspended in a fluid as resulting from fluid molecules in thermal motion acting on the particle. This diffusion is typically characterised through the slope of the particle's mean-squared displacement,
\begin{equation}
    \Dself(t) = \frac{1}{2d}\dv{t} \left\langle  \left|\vb{r}(t + t_0) - \vb{r}(t_0) \right|^2 \right\rangle
    \label{eq:msd}
\end{equation}
where $\vb{r}(t)$ is the particle position at time $t$ in $d$ dimensions and $\langle \cdot \rangle$ indicates averaging over all starting times $t_0$. 
At infinite particle dilution, $\Dself(t)$ provides the \textit{bulk} or \textit{free} diffusion coefficient $ \Dself(t) \equiv \Dbulk = k_B T/\gamma$, where $k_B T$ is the unit of energy and $\gamma$ a friction coefficient (Fig.~\ref{fig:intro}-a).
When a particle diffuses in a suspension of particles, its \textit{self} diffusion coefficient $\Dself(t)$ may differ from $\Dbulk$ (Fig.\ref{fig:intro}-b). It may be reduced due to interactions with neighbors, \textit{e.g.} through a caging effect in the presence of repulsive interactions, or via hydrodynamic interactions~\cite{dhont1996introduction}. In general, it depends on time. We define its short-time value, $\Dself = \Dself(t \simeq 0)$. %\adam{is it needed to get rid of the $t$ here?} \sophie{yes for clarity further?}%\alice{I feel like you already said quite a bit of this above right?}

When a density gradient forms in a particle suspension, the so-called \textit{collective} diffusion coefficient $\Dcoll$ characterises the relaxation of the gradient, which is inherently a many particle behaviour (Fig.~\ref{fig:intro}-c). Note that this coefficient is sometimes referred to as the transport diffusion coefficient, and we refer the reader to Ref.~\cite{schlaich2024theory} for an overview.
Most often, collective diffusion is defined via a \textit{wavevector-dependent} diffusion coefficient $D(k, t)$ which characterizes the relaxation of a density fluctuation $\delta \rho = \rho(k, t) - \langle\rho\rangle$ at a wavelength $k$ in a colloidal suspension as
\begin{equation}
    \delta \rho(k, t) = \delta\rho(k, t=0) e^{-D(k, t)k^2t}.
\end{equation}
In the large wavevector (short length scale) limit, $D(k,t)$ recovers the self diffusion coefficient, $\Dlargekt = \Dself(t)$. Conversely, in the vanishing wavevector (large length scale) limit, $D(k,t)$ gives the collective diffusion coefficient, $\Dsmallkt = \Dcoll(t)$. In a few cases, \textit{e.g.} for small density fluctuations or for pairwise additive interactions, the collective diffusion coefficient is independent of time ~\cite{dhont1996introduction, pusey_liquids_1991}. %\adam{haven't really defined long or short time}.

%Collective diffusion is useful to characterize the macroscopic transport properties of a system. %, whether to assess the relaxation speed of complex suspensions or test the permeability of porous matrices to various molecules~\cite{sobac2020collective,ariskina2022free,obliger2016free}. However, 
In contrast to self-diffusion, determining how individual interparticle interactions govern collective diffusion is complex. 
%For example, in a suspension with hard-sphere interactions (and significantly far from the glass transition), self-diffusion is hindered by neighbouring particles inducing spatial constriction via the caging effect. In contrast,  
In general, collective diffusion is enhanced by interparticle interactions for hard-spheres, as moving particles push their neighbours, facilitating relaxation. A theoretical argument considering only pairwise %\adam{what does pair-wise mean here? Can we replace with ``hard sphere''?} \sophie{Hard sphere is a type of pair wise interactions, but this formula is valid for soft interactions, not just hard sphere. so pair wise is the suitable word. Pair wise means its only 2 by 2 interactions. In contrast, N-body interactions would mean that the interaction potential depends on the relative position of all or many of them at the same time. } 
interactions shows $\Dcoll = \Dself/\SkZ$, where $\SkZ$ is the structure factor of the suspension at vanishing wavevector~\cite{gomer1990diffusion,dhont1996introduction}. %\adam{I have been unable to find this in \cite{gomer1990diffusion}. Pusey Les Houches (5.84) is an unsatisfying derivation but says nothing about ``pair-wise''.} \sophie{Eq. 2.26 and then 2.91 give this formula, up to a density prefactor which depending on the definition of $S(k=0)$ should or should not be there (in our case it isn't).}. 
At high particle densities in colloidal suspensions, $\SkZ$ can be substantially smaller than 1, and therefore $\Dcoll$ may indeed be significantly larger than $\Dbulk$. This yields $\Dself \leq \Dbulk \leq \Dcoll$ \cite{dhont1996introduction},  highlighting how subtle collective diffusion is, even in a simple case. %\adam{I am not sure about this sentence. $\SkZ < 1 \Rightarrow \Dcoll > \Dbulk$ doesn't seem especially subtle}. 
For hard sphere-like colloids in a suspension, an added complexity in verifying these theoretical models for how steric interactions influence collective diffusion is the inescapable presence of hydrodynamic interactions. In contrast to the case of self-diffusion, the effect of hydrodynamic interactions on collective diffusion coefficients is not well established and experimental results widely differ in assessing the role of hydrodynamic interactions on collective properties~\cite{panzuela2018solvent, panzuela2017collective,kops1982dynamic,qiu1990hydrodynamic, segre1995experimental, bleibel20153d, lin1995experimental,lin2014divergence,bleibel2014hydrodynamic,bleibel20153d, falck2004influence,mackay2024countoscope}. %\adam{do we need to measure hydrodynamic interactions here? it reads as a little ``out of the blue to me''} \alice{I agree - I understand we are trying to move away from the hydro stuff but I guess we can't ignore it completely} 
Methods to reliably measure $\Dcoll$ are thus needed to shed light on collective motion.

%\sophie{I moved something on hydrodynamic interactions to the discussion since the intro is long enough and we do not resolve this question}
% n addition to direct interactions set by the interparticle potential, the suspending fluid also introduces hydrodynamic interactions - long range particle-particle interactions mediated by the fluid. %\sophie{text}
% %Particle interactions are generally more complex than simply repulsive and include at least interactions mediated by the suspending fluid, or hydrodynamic interactions. 
% However, even for suspensions of hard sphere-like colloids, experimental results widely differ in assessing the role of hydrodynamic interactions on collective properties~\cite{panzuela2018solvent, panzuela2017collective,kops1982dynamic,qiu1990hydrodynamic, segre1995experimental, bleibel20153d, lin1995experimental,lin2014divergence,bleibel2014hydrodynamic,bleibel20153d, falck2004influence,mackay2024countoscope}. 
%Such discrepancies arise from the challenge of 
Measuring collective diffusion coefficients from particle coordinates, both experimentally and numerically, is challenging. Since collective diffusion manifests out-of-equilibrium, several investigations explore the relaxation of a number density gradient~\cite{loussert2016drying,keita2021microfluidic,sobac2020collective,merlin2011time}. However, such experiments are hard to repeat, because they require setting up the system out-of-equilibrium at the beginning of each experiment. In addition, as the density gradient relaxes, neither the gradient nor the local density is constant, which makes it hard to disentangle how $\Dcoll$ depends on particle number density. Collective diffusion can also be probed in equilibrium, from density fluctuations that occur due to thermal motion. 
The relaxation of the density fluctuations is then investigated through the dynamic structure factor calculated in Fourier space~\cite{dhont1996introduction}. Yet, Fourier transforms are computationally demanding and fraught with spurious features due to edge effects on microscopy images~\cite{moisan2011periodic,giavazzi2017image,zuccolotto2024improving}. At equilibrium another strategy consists in probing the diffusion coefficient of the centre of mass of many particles~\cite{obliger2016free,ariskina2022free,schlaich2024theory,ala2002collective,falk2015subcontinuum,thorneywork2015structure}. %$D_{\mathrm{cm}}$ is sometimes referred to as ``collective'' diffusion coefficient, in which case $\Dcoll$ is termed ``transport'' diffusion coefficient. 
The collective diffusion coefficient is then proportional to the diffusion coefficient of the centre of mass~\cite{gomer1990diffusion}. %\adam{I have just noticed that this implies $D_{\mathrm{cm}} = \Dself$ which seems wrong}. 
Yet, obtaining a statistically meaningful trajectory for the centre of mass requires following a large group of particles for a substantial amount of time, which is experimentally challenging as particles continuously exit and enter the field of view. Even in simulations, only one trajectory is obtained, limiting statistical resolution.

Here, we establish a novel approach to measure collective diffusion coefficients experimentally and numerically by investigating the statistics of particle number counts $N(t)$ in virtual observation boxes at equilibrium (Fig.~\ref{fig:intro}-d). The number $N(t)$ fluctuates due to particles diffusively entering and exiting a box (Fig.~\ref{fig:intro}-e). For small observation boxes, fluctuations are dominated by individual particle motion. Large observation boxes sense collective motion since fluctuations also reflect the relaxation of transient groups of particles. In both cases, relaxation of fluctuations is linked to a characteristic time scale $T(L)$, which shows a non-trivial dependence on the size of the square box $L$. From this time scale, a length scale dependent diffusion coefficient has been defined, $D(L) \sim L^2/T(L)$, which is sensitive to $D_{\rm self}$ for small length scales, $D(L\rightarrow 0) = D_{\rm self}$, and to collective diffusion at large length scales $D(L \rightarrow\infty) = \Dcoll$. This idea, termed the ``Countoscope'' (Sec.~\ref{sec:countoscope intro}), was proposed recently by some of us to probe self diffusion, yet its potential to probe collective diffusion coefficients %, the behaviour of D(L) for large length scales,  
has not been conclusively explored~\cite{mackay2024countoscope}. %For example, among collective properties, are fluctuating counts indeed sensitive to $\Dcoll$? Can we build a workflow and understand best practices to extract $\Dcoll$ from experimental data? And how do fluctuating counts compare with existing methods to measure collective properties?
%In this work, we investigate the equilibrium diffusive dynamics of 2D colloidal suspensions experimentally and numerically with fluctuating counts. We demonstrate that counting is indeed sensitive to $\Dcoll$ in the limit of large box sizes. Beyond $\Dcoll$, counting can illuminate the relaxation dynamics at all lengthscales via a box-size dependent diffusion coefficient $D(L)$ (where $L$ is the size of the square box), in analogy with $D(k)$. We establish a workflow to measure $D(L)$ and hence $\Dcoll$ by obtaining the relaxation timescale of fluctuations via data integration. 
In this work, we determine best analysis practices and discuss perspectives to optimize this measurement (Sec. \ref{sec:countoscope to measure collective diffusion properties}). 
We compare our results with dynamic structure factor approaches, and find Fourier techniques yield inaccurate long-range collective measurements because of the non-periodic nature of an experimental image. In contrast, counting exploits finite images by paving the space with finite virtual observation boxes (Sec.~\ref{sec:comparison to fourier}). Finally, we discuss how our methodology could provide further insights into the collective properties of suspensions%. , particularly in understanding the role of hydrodynamic interactions 
(Sec. \ref{sec:discussion and conclusion}).

% \begin{itemize}
%     \item collective d generally useful to understand macroscopic transport properties e.g. in porous media~\cite{ariskina2022free,obliger2016free}
%     \item \cite{ariskina2022free} has a nice discussion on why self = center of mass in the absence of long-range hydrodynamic interactions. 
%     \item \cite{panzuela2018solvent} very good paper by Rafa which shows long and short collective d feel hydrodynamics in a very different way (cf membrane geometry)
%     \item \cite{panzuela2017collective} simulation 2D bulk shows divergence at small k, and proposes an interesting rescaling
%     \item \cite{kops1982dynamic} 3D experimental data, very scarce k wave numbers, shows divergence but only over a very short range... 
%     \item Adam mentions that the Bleibel paper says the out-of-plane of motion might be killing the divergence; and also the Lin paper says the mechanism for hydrodynamic enhancement is different in the 2 2D cases. \adam{I no longer think the out of plane motion kills the divergence - actually maybe it just pushes it to bigger length scales. They say something about how if you keep zooming out eventually $L >>> \text{out of plane}$ and so at some point it will stop mattering} 
% \end{itemize}

\section{Brief introduction to the Countoscope with overlapped boxes}
\label{sec:countoscope intro}

%\adam{Sophie is there a reason you put the countoscope before $f(k, t)$? I would have put $f(k, t)$ first as it's the pre-existing method} \sophie{Yes, as in most important/newest things/focus of the paper first. If the story on $f(k,t)$ were simple we could put it first but it's not really...} \sophie{But I'm open for discussion.}

\subsection{Experimental system: 2D colloidal suspension near a wall.}
We investigate the collective diffusive relaxation of a 2D suspension of colloids experimentally and numerically. We briefly restate the system's properties here and refer to Ref.~\cite{mackay2024countoscope} for details. 

\begin{figure}[h!]
    \centering
    \includegraphics[width=\linewidth]{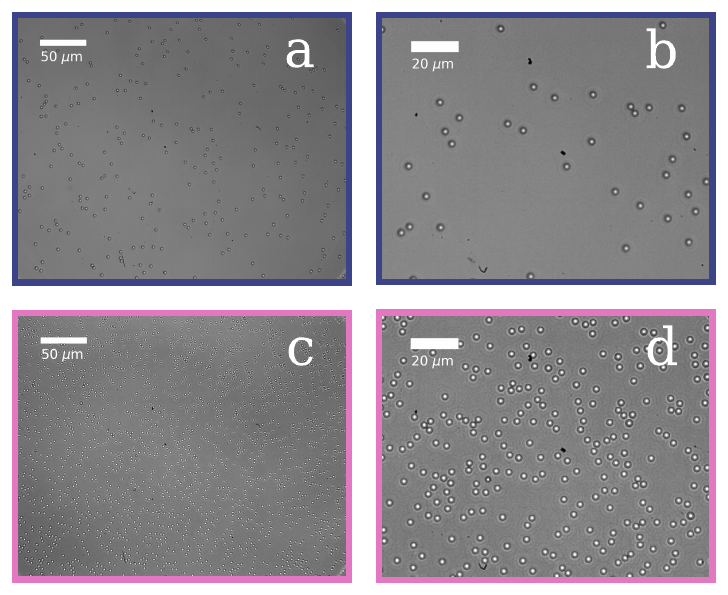}
    \caption{\textbf{2D experimental hard-sphere system.} Optical images of (a,b, dark blue) the dilute $\phi = 0.02$ and (c,d, pink) the dense $\phi = 0.11$ suspensions. (a,c) shows the entire field of view and (b,d) a cropped subset for visualization. %\sophie{Tune saturation level to make similar gray scale, and less white space.}
    }
    \label{fig:experiments}
\end{figure}

In experiments, particles, with effective hard sphere diameter $\sigma = \SI{3.0}{\micro\meter}$, which represent well a hard sphere model~\cite{thorneywork2014communication,Thorneywork2015}, are gravitationally confined in the $z$-direction to the base of a glass cell. Suspensions are imaged in the $x$-$y$ plane using a custom-built inverted microscope -- see SI section \ref{sec:experiments} for more details. Data is recorded at 2 fps and the total field of view is approximately $\qtyproduct{290 x 360}{\micro\meter}$. Particle positions are acquired from images using standard particle tracking protocols~\cite{crocker1996methods,allan_daniel_b_2021_4682814}. For simplicity, we explore here only two different packing fractions, corresponding to a dilute ($\phi = 0.02$) and dense ($\phi=0.11$) suspension -- see Fig.~\ref{fig:experiments}. Note that while $\phi=0.11$ lies far below the crystallisation transition for these systems, this packing fraction is sufficiently dense that interactions play an appreciable role. %Importantly, however, these interactions do not result in such long timescales that the full dynamic properties are not accessible over reasonable experimental timescales. \adam{would a higher packing fraction result in longer timescales? I don't think so?}.\alice{ if you look at the PRX, I think it takes longer for the theory curve to plateau as $\phi$ goes up. It is also generally true that as you approach a phase transition, correlation length scales and time scales increase significantly, but potentially not all timescales are longer. You probably hit the long-time diffusion coefficient faster because direct interactions become more significant more quickly. I think we need an argument for why we only go up to 0.11, but it could be more vague - we certainly avoid complexities associated with higher phi this way} \sophie{I think the way its written is fine as it is.} 
To obtain statistically accurate data over the long collective relaxation timescales we investigate, we acquire experimental data over $\SI{20}{h}$. The optical stage and experimental conditions were carefully adjusted to avoid any significant drift over this long time scale (see SI section 1.1). %\adam{is this what the reviewer was talking about/is this repeated in the SI?}.
At these packing fractions, the short-time self diffusion coefficient of particles can be calculated through Eq.~\eqref{eq:msd} averaged over many particles, and gives $\Dself = \SI{0.048 \pm 0.001}{\micro\meter^2\per\second}$ %\adam{this is not compatible with the PRX value (0.043)} 
at $\phi = 0.02$ and $\Dself =  \SI{0.043 \pm 0.001}{\micro\meter^2\per\second}$ at $\phi = 0.11$.

In parallel, we conduct Brownian dynamics simulations representative of the experimental system. Simulation parameters are all set to their experimentally measured values, and steric forces are modelled using the hard potential in Eq.~(31) of Ref.~\cite{sprinkle2020driven}. Hydrodynamic interactions between particles  -- both far field and lubrication interactions -- are not accounted for in the simulation for simplicity. This allows for a direct comparison between the collective diffusion coefficient obtained from our simulations and that predicted by theory, which does not include hydrodynamics, to demonstrate the capabilities of our method. For the experiments, lubrication (near-field) interactions are not expected to play a significant role, as the mean distance between particles is large, greater than $3\sigma$ at the largest packing fraction $\phi = 0.11$. Far-field hydrodynamics do play a role as seen in the reduction of the short-time self-diffusion coefficient $\Dself$ at $\phi = 0.11$. %\adam{I thought they didn't play much of a role because of screening?}.\alice{screening by the bottom wall does not seem to be as significant as people claim, I think two walls would have more effect}. 
As $D(k)$ and $\Dcoll$ are expected to linearly depend on $\Dself$ at first order in the hydrodynamic interactions, we account for this by presenting most results renormalized by $\Dself$. %Finally, given that our goal is to demonstrate the capabilities of the method rather than disentangle the role of hydrodynamics, we keep simulations without hydrodynamic interactions between particles for simplicity.
Finally, hydrodynamic lubrication between particles and the bottom glass wall is accounted for~\cite{sprinkle2020driven} through the value of $D_0$. %\alice{or also through the value of $\Dself$}. 
This is reduced with respect to the value for a particle in the bulk by a factor consistent with theoretical predictions~\cite{goldman1967slow,sprinkle2020driven}.  
%These interactions reduce the self diffusion coefficient $\Dself$, and as $D(k)$ and $\Dcoll$ are expected to linearly depend on $\Dself$, they will not affect the renormalised quantities that we will present, $D(k)/\Dself$ and $\Dcoll/\Dself$. 
Further numerical and simulation methods are described in SI section \ref{sec:simulations}.

\vspace{3mm}

\subsection{Number fluctuations in observation boxes.}
For both experiments and simulations, we sample fluctuations 
in the number of particles $N(t)$ within square boxes of size $\Lbox\times \Lbox$ over time. 
$N(t)$ fluctuates between discrete values as a consequence of particles moving in and out of the box via diffusion (Fig.~\ref{fig:intro}-e). 
We explore the statistical properties of this random number $N(t)$. We can compute the correlation function depending on the lag time $t$ as
\begin{equation}
C_N(t) = \brak{N(t+t_0) N(t_0)} - \Nmean^2
\end{equation} where $\langle \cdot \rangle$ indicates an average over all boxes and time origins $t_0$ within the acquisition. For simplicity in the following we write $t_0 = 0$. Notice that when the lag time vanishes, $C_N(0) = \langle N^2 \rangle - \langle N\rangle^2 \equiv \Nvar$.
Another relevant quantity is the mean squared change in particle number, 
\begin{equation}
\begin{split}
    \DNtt  &= \brak{\left(N(t) - N(0)\right)^2} \\
    &= 2 (\brak{N^2} - \Nmean^2) - 2 (\brak{N(t) N(0)} - \Nmean^2) \\
   &=  2\Nvar - 2 C_N(t).
\end{split}
\label{eq:nMSD}
\end{equation}

\begin{figure}
    \centering
\includegraphics[width=0.99\linewidth]{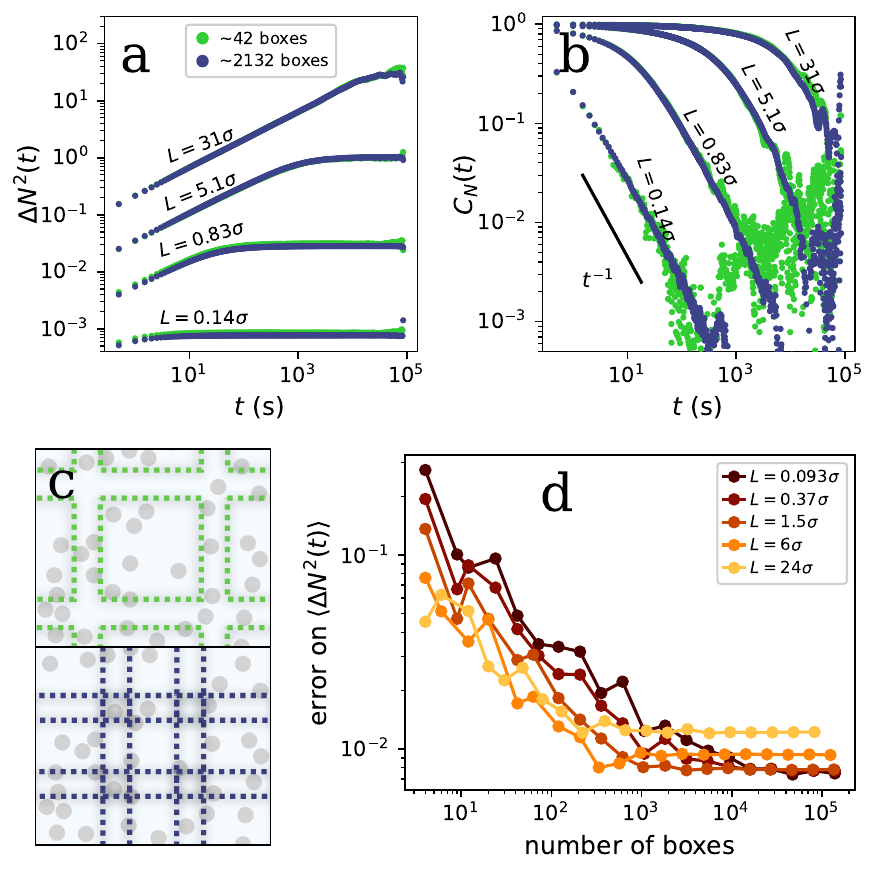}
    \caption{\textbf{Importance of overlapping observation boxes.} (a) Number fluctuations $\DNtt$ and (b) correlation function $C_N(t)$ as a function of the lag time $t$ for several box sizes for the dilute suspension $\phi = 0.02$. Experimental data. Legend is shared between a and b. (c) Schematic illustrating box overlapping. (d) Normalised average standard error on $\DNtt$ with respect to the number of observation boxes for different box sizes. The error is calculated by dividing the experimental data at $\phi=0.02$ into 10 chunks in time, computing $\DNtt$ for each chunk, and finding the (averaged over time) standard error between chunks.
    }
    \label{fig:overlapped}
\end{figure}

%$\DNtt$ is analogous to the mean-squared displacement. \adam{should we make this comment? I find it somewhat unhelpful because the NSMD doesn't really behave the same way} 
Both statistical quantities will be useful to investigate as they characterize the dynamical relaxation of number fluctuations. 
In Fig.~\ref{fig:overlapped}-a, we plot the mean squared change in particle number, $\DNtt$, for different box sizes in the dilute regime ($\phi = 0.02$). $\DNtt$ first increases in time. Starting from an initial condition with $N(0)$ particles in a box, as time goes by, one is more and more likely to see configurations where $N(t)$ is much higher or much smaller than $N(0)$, resulting in an overall increase of the squared difference $(N(t)-N(0))^2$ on average. Eventually, there is significant exchange between particles inside the box with those outside and we observe a plateau. The number of particles at long times is therefore uncorrelated with that in the initial configuration, \textit{i.e.} $C_N(t) \simeq 0$ for long times (Fig.~\ref{fig:overlapped}-b). 
%The plateau can be understood as corresponding to complete exchange of particles inside the box with those outside. At this point, the number of particles is uncorrelated with the initial configuration, $C_N(t) \simeq 0$. 
From Eq.~\eqref{eq:nMSD}, the plateau corresponds to the variance $\langle \Delta N (t \rightarrow \infty) \rangle = \ 2 \Nvar$.  

In Ref.~\cite{mackay2024countoscope}, we established that number fluctuations can resolve the short-time self-diffusion coefficient of particles $\Dself$. Indeed, at short times, fluctuations are dominated by individual particles entering or exiting boxes and satisfy ${\DNtt \sim \sqrt{\Dself t/L^2}}$. In a system with no interactions, the fluctuations relax over a timescale of about $L^2/\Dself$. Yet for denser suspensions, over longer timescales, and especially in large boxes, the relative motion of groups of particles, or collective dynamics, should affect number fluctuations.

\vspace{3mm}

\subsection{Overlapping boxes.}
In this work, we aim to characterise the relaxation of number fluctuations, especially at large length scales -- in large boxes -- where collective dynamics are at play.  
Obtaining statistically accurate data over large boxes is inherently challenging, but can be optimized by carefully choosing how to distribute boxes spatially. %Obviously, if one keeps observation boxes separated (Fig.~\ref{fig:overlapped}-c, green case), then the number of large boxes one can fit on a single image is much smaller than for small boxes. 
With separated boxes (Fig.~\ref{fig:overlapped}-c, green case), as in Ref.~\cite{mackay2024countoscope}, statistical information is greatly reduced at large scales as fewer boxes fit on an image, with significant unused regions on an image. Instead, we propose to overlap sampling boxes (Fig.~\ref{fig:overlapped}-c, blue case), significantly increasing the number of observation boxes. Qualitatively, overlapping boxes improve the resolution of the plateau of $\DNtt$ (Fig.~\ref{fig:overlapped}-a), and significantly reduce noise on the long tails of the correlation function $C_N(t)$ (Fig.~\ref{fig:overlapped}-b). This means that, although one obtains somewhat correlated data with overlapped boxes, the increased sampling is more important and improves statistical resolution. 

To understand what degree of overlapped boxes yields the best statistics, we evaluate the average standard error on $\DNtt$ as we increase the number of boxes, and hence the amount of overlap between boxes (Fig.~\ref{fig:overlapped}-d). The error decreases by an order of magnitude with increasing box numbers, confirming overlapped boxes significantly improve statistical accuracy. Eventually the error reaches a noise floor: as we pave space with boxes, boxes eventually become so overlapped that they are redundant and no more information is gained. Since this excessive overlap clearly happens with fewer boxes for larger boxes, the noise floor is reached with fewer boxes for large boxes than small ones. 
Overall, this suggests an upper limit for the number of boxes to use, which we take here to be 2000 boxes for our system parameters. We use this overlapping technique and bound in all future analysis. 

%%%%%%%%%%%% METHOD 2

\section{Countoscope to measure collective diffusion properties}
\label{sec:countoscope to measure collective diffusion properties}

We now use the Countoscope to investigate the relaxation of number fluctuations, focusing on how this relates to collective diffusion properties for large boxes. We do this by defining a box-size dependent coefficient $D(L)$, in analogy with $D(k)$ as discussed earlier. The two statistical quantities, $\DNtt$ and $C_N(t)$, allow us to explore two complementary methods to extract collective dynamics. In this section, we find $D(L)$ from a relaxation timescale $T(L)$, obtained by integrating the correlation function $C_N(t)$. We term this method the ``timescale integral''. In the SI (section \ref{sec:alternative method}) we present an alternative method, using a phenomenological fit of the number fluctuations $\DNtt$ to obtain $D(L)$. This second method is more suited to situations where the dataset length is limited, but it requires a phenomenological model of the effect at play, and is less accurate for sufficiently long datasets. 
We will compare results between an effectively non-interacting case, the dilute suspension at $\phi = 0.02$, and the denser suspension at $\phi=0.11$ where interactions modify behaviour. We will also explore differences between experimental data, simulations and theory. 
Importantly, we will show $D(L)$ interpolates between two regimes, a regime in which self-diffusion dominates in small boxes (red in Fig.~\ref{fig:intro}-d) and collective diffusion in large boxes (orange in Fig.~\ref{fig:intro}-d).

%\subsection{Decorrelation timescale of number fluctuations}
%\label{sec:timescale integral}

% \vspace{2mm}

\begin{figure}[h!]
    % \centering
    \includegraphics[width=0.99\linewidth]{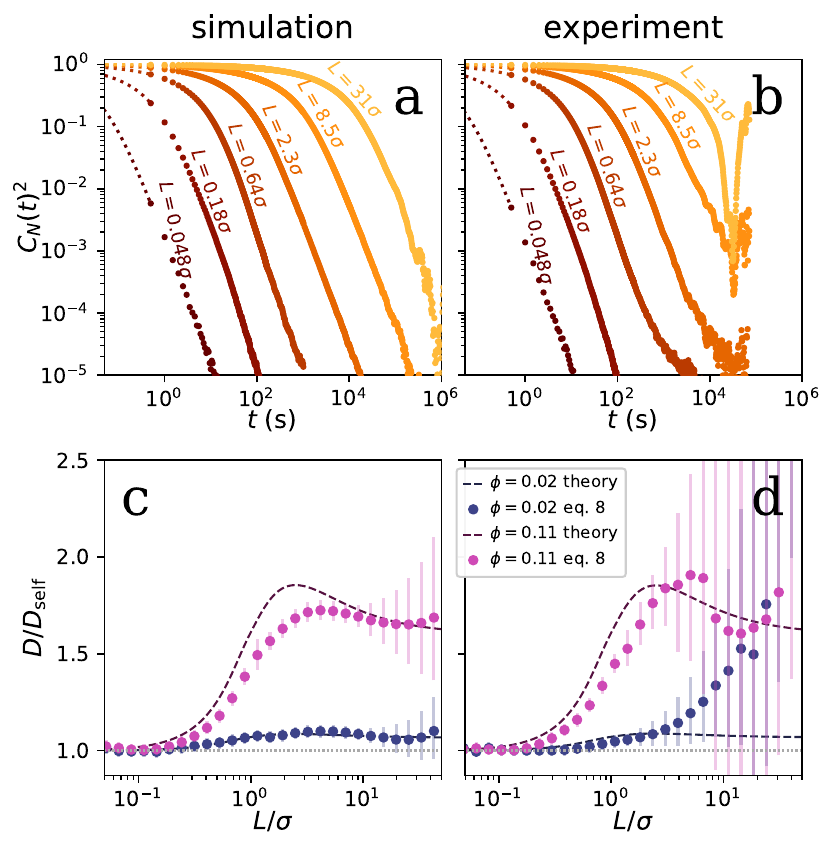}
    \caption{
        \textbf{Decorrelation timescale of number fluctuations.} 
       (a-b) Correlation functions  $C_N(t)^2$ versus time for (a) simulations and (b) experimental data in the dense $\phi = 0.11$ case. Box sizes go from small (dark red) to large (yellow). Dotted curves for $t \leq 0.5$~s represent short-time extensions to improve the accuracy of the integral of $C_N(t)^2$ on small boxes $L \leq \sigma$. (c-d) Diffusion coefficient $D(L) \propto L^2/T(L)$, where $T(L)$ is obtained from the integral of $C_N(t)^2$ from (a-b) for (c) simulations and (d) experimental data. Dashed lines correspond to the theory Eq.~\eqref{eq:time} (using also Eq.~\eqref{eq:Cnt}). Blue corresponds to the dilute regime, and pink the dense. %\sophie{Waiting for updated curves. The simulation should nicely agree (with a small error) with the theory. The experimental one should not diverge as much. We may or may not include the long time fits.}
       Error bars are propagated from 99\% confidence intervals in the variance of $N(t)$ across all boxes and times. 
    }
    \label{fig:timescale}
\end{figure}

\subsection{Workflow of timescale integral method.}
\label{sec:timescale integral}

%\vspace{3mm}

The timescale integral method is based on quantifying the relaxation time of the correlation function $C_N(t)$ (see Fig.~\ref{fig:timescale}-a and b). 
%As particles diffusively enter and exit the box, the correlation function decays as the number of particles in the box becomes increasingly decorrelated with its previous value. 
To quantify the timescale $T(L)$ of this decay, Ref.~\cite{mackay2024countoscope} suggested to integrate $C_N(t)$ as 
\begin{equation}
\begin{split}
    T(\Lbox) &= 2 \int_0^{\infty}  \left(\frac{C_N(t)}{C_N(0)} \right)^2 \, \dd t \\
    &=  2 \int_0^{\infty}  \left(1 - \frac{1}{2}\frac{\DNtt}{\Nvar} \right)^2 \, \dd t. 
       \label{eq:time}
\end{split}
\end{equation}
%This is the dynamic analogue of the number variance in Eq.~\eqref{eq:plateau}. 
This definition means that for correlations that decay exponentially, $T$ would represent the timescale of the decay as $C_N(t)/C_N(0) = \exp(-t/T)$. Note, here the unusual squaring factor in the integrand ensures that the integral converges, as the correlation function decays algebraically at long times, $C_N(t\rightarrow \infty) \sim 1/t$~\cite{mackay2024countoscope,mackay2024noise} (Fig.~\ref{fig:overlapped}-b). Visually, $T(L)$ also roughly corresponds to the corner in the $\DNtt$ curves.

\subsection{Scale-dependent diffusion coefficient $D(L)$ can probe $\Dcoll$}
In the dilute regime, one can verify experimentally that $T$ corresponds to the time to diffuse across the box, $T(\Lbox) \propto  \Lbox^2/4\Dbulk$;  %where $\alpha_T \simeq 0.56$ is a numerical constant which can be obtained from theory and whose lengthy expression is reported in Eq.~\eqref{eq:alphaT}. 
for dilute or non-interacting systems, rescaling time by $L^2$ is sufficient to describe the relaxation of fluctuations at all scales. At higher packing fractions, the time required to relax fluctuations does not solely depend on how long a single particle takes to diffuse over the length scale $L$ (SI fig \ref{fig:nmsd rescaled}). Instead, it depends on the motion of multiple interacting particles.
%(SI fig \ref{fig:nmsd rescaled}-a). a simple rescaling of the data fails (SI fig. \ref{fig:nmsd rescaled}-b), indicating that %\adam{can you read this paragraph without reading the appendix?}

We can relate $T$ to a general diffusive phenomenon, by defining a diffusion coefficient dependent on the box size, as 
\begin{equation}
    D(\Lbox) = \alpha_T \frac{\Lbox^2}{4 T(\Lbox)}
    \label{eq:Dbox}
\end{equation} 
where $\alpha_T \simeq 0.56$ is a numerical constant, which can be obtained from a theory we describe below and whose lengthy expression is reported in Eq.~\eqref{eq:alphaT}. 
We present $D(L)$ rescaled by $\Dself$ in Fig.~\ref{fig:timescale}-c and d for the 2 different packing fractions in this study, for simulations and experiments.

To gain insight on the behaviour of $D(\Lbox)$ we first investigate it with theory. Predictions for $C_N(t)$ can be obtained by describing the particle suspension with a stochastic density field theory (Dean-Kawasaki equations)~\cite{dean1996langevin,kawasaki1994stochastic,mackay2024countoscope}. In this theory, particle interactions are pair-wise interactions characterized via their structure factor $S(k)$. Thus, the theory is general in that it applies to any particle suspension with pair interactions. Here, we only include hard sphere interactions, thereby reproducing the ingredients of the simulations, and the expression of $S(k)$ is given by Eq.~\eqref{eq:Sk}. The theory is described in Ref.~\cite{mackay2024countoscope} and recapitulated in SI section \ref{sec:counts in large boxes}. 
In the analytical curves (lines in Fig.~\ref{fig:timescale}-c and d), $D(L)$ plateaus for small boxes, increases, and eventually plateaus again. It can be shown analytically that for small boxes the timescale integral probes self diffusion,  and for large boxes it probes collective diffusion 
\begin{equation}
    \begin{cases}
        &\displaystyle \DsmallL = \Dself, \\
        &\displaystyle \DlargeL = \Dcoll = \frac{\Dself}{\SkZ}.
    \end{cases}
\end{equation}

%\vspace{3mm}

We check the theory predictions by investigating $D(L)$ with our simulations and experiments.
For both packing fractions, simulations closely follow the theory over nearly the full range of box sizes: $D(L)$ plateaus for small boxes, increases, and then appears to plateau again. Experimental results also follow the theory up to around $L \gtrsim \sigma $ before hitting a divergent region. %\adam{need to be clearer what we're talking about here, the simulations do not diverge at $L=\sigma$}.\alice{this would be my take on it here} 
This divergence is due to limited statistical accuracy, and we discuss it further in Sec.~\ref{sec: countoscope accuracy}. For small boxes $L \lesssim \sigma$, we find ${D(L) \simeq D_{\rm self}}$ as expected, the box-size dependent diffusion coefficient probes individual motion. Collective motion then increases the effective diffusion coefficient $D(L)$ at intermediate lengthscales.
In the simulations, $D(L)$ clearly reaches a plateau around ${L \simeq 10 \sigma}$, which corresponds closely with the expected value of the collective diffusion coefficient $\Dcoll$ of the suspension. This plateau is also apparent, though slightly less so in experimental data, and we discuss how to improve the measurement in Sec.~\ref{sec: countoscope accuracy}. Overall, the timescale integral can indeed capture $\Dcoll$ at large lengthscales.

At intermediate lengthscales, $L \simeq 2-4\sigma$, some discrepancies can be noticed between simulations, experiments and theory, for the largest packing fraction $\phi = 0.11$. As they do not appear for $\phi = 0.02$ these discrepancies arise because of particle interactions. In the theory, particle interactions are linearized assuming density fluctuations are small. From the difference between theory and simulations -- which contain, as the theory, only hard sphere interactions -- we can conclude that large density fluctuations arise at these intermediate lengthscales. Further discrepancies are observed between experimental data and simulations, which are likely arising from far-field hydrodynamic interactions. 

Finally, experiments, simulations and theory highlight the presence of a peculiar maximum in the $D(L)$ curves in the dense case $\phi = 0.11$ (pink), near ${L \simeq 3 \sigma}$. %This maximum could be present in experimental data as well, as can be seen, if we attempt to improve large box accuracy (see SI Fig). 
This effect is quite subtle, and here we propose a qualitative explanation. For small box sizes, ${L \ll \sigma}$, one probes individual motion at small lengthscales, corresponding to a timescale much smaller than the mean time between collisions. For large box sizes, ${L \gg \sigma}$, one probes collective motion, \textit{i.e.} the relaxation of transiently forming groups, at scales much larger than these transient groups, in essence at a mean-field level. For intermediate yet small box sizes, say ${L \simeq \sigma}$, one still probes individual motion but at a scale where a particle senses its neighbours. The interacting neighbours facilitate relaxation of number fluctuations, by pushing one another. This results in an increase of effective dynamics $D(L)$. This increased $D(L)$ can exceed $\Dcoll$, because the box is still small enough, $L \simeq \sigma$, that the magnitude of the number fluctuations that relax is still small on average. Schematically, at these scales one still investigates only individual particles that get effectively pushed out by their neighbours, resulting in a maximum of $D(L)$. We will comment on such behaviour further in Ref.~\cite{mackay2024noise}. 

\subsection{Best practices for obtaining the timescale integral.}
\label{sec: countoscope accuracy}

To perform the timescale integral accurately one must obtain an accurate estimate of the correlation function $C_N(t)$. Here, a significant challenge is to accurately calculate $\Nvar$. To obtain $C_N(t)$, the mean squared change in particle number $\DNtt$ is subtracted from its plateau, the variance $\Nvar$, as, according to Eq.~\eqref{eq:nMSD}, $C_N(t) = \left[\DNtt/2 - \Nvar \right]$. Inaccurate estimates of $\Nvar$ thus result in divergences in $T(L)$, and on $D(L)$, as observed for large enough boxes %${L\gtrsim 20 \sigma}$ 
in Fig.~\ref{fig:timescale}-c and d. The fact that $C_N(t)$ does not vanish smoothly for the boxes ${L \geq 2 \sigma}$ in Fig.~\ref{fig:timescale}-b, also demonstrates that $\Nvar$ is not accurately resolved. 

In general, 2D systems require long times to sample many different states, such that accurate estimates of $\Nvar$ are hard to access. A long-time scaling law can be obtained from the theory showing that $C_N(t)$ decorrelates slowly, as $1/t$ (see Fig.~\ref{fig:overlapped}-b). This is a consequence of the fact that diffusing particles can always return to their starting point in 2D~\cite{mackay2024noise}. Resolving $\Nvar$ in 2D thus requires significantly long datasets, especially for larger boxes.
Due to the slow decay of number correlations, we expect experiments need to be at least as long as the decorrelation time to resolve $\Nvar$ correctly. Indeed, for $L = 20 \sigma$, the decorrelation time is roughly $T \simeq L^2/\Dbulk \simeq \SI{23}{h}$. Since this is about the duration of our experiments, this justifies that divergences appear in Fig.~\ref{fig:timescale}-d for $L \gtrsim 20 \sigma$. Generally, this defines an upper box size that can be resolved in an experiment, as $L \lesssim \sqrt{\Dbulk T_{\rm exp}}$ where $T_{\rm exp}$ is the experiment time. %\alice{to come back to your question about timescales earlier - if $D_0$ here is the self diffusion coefficient, this will be lower at higher packing fraction and so these timescales will be even longer, which justifies us working with the lowest packing fraction system that shows up the interactions }

Simulations without hydrodynamics can be conducted for long enough times that convergence in time is not an issue. Yet, we find reliable variance estimates are also hard to access for large boxes. %the largest boxes presented, $L \gtrsim 40 \sigma$. 
We identify that better estimates of $\Nvar$ may be obtained by increasing the size of the periodic simulation box $L_x$ (see SI Fig \ref{fig:plateaus sim rescaled}). Indeed, collective dynamics loop back onto themselves via the simulation's periodic boundaries, and hence large simulation boxes are needed for accurate resolution. This defines an upper box size that can be resolved in a simulation, as  $L \simeq 0.3 L_x$. Here we have $L_x = \SI{1280}{\micro\meter}$ and so on boxes smaller than $L \lesssim 130 \sigma$, accurate resolution is possible, which is apparent in Fig.~\ref{fig:timescale}-c. %\adam{with the new bigger simulations, $0.3L_x$ has changed and is now way off the plot}. 
Finally, we note that, in this work, we take the variance as an average over all boxes and all times. To estimate the variance from such time-correlated data, other strategies such as bootstrapping could be used~\cite{efron1992bootstrap}, or fitting the distribution of particles in a box. However, we find that our dominant source of error is not the method to estimate the variance but rather (\textit{i}) lack of long time data in experiments, or (\textit{ii}) lack of large simulation box in simulations.

% Missing data at short and long timescales
The length of the dataset is also important in that, to integrate $C_N(t)$, one must have data over long enough time periods. 
During integration, the decay at long times could be in part improved by replacing the long-time noisy data points with long-time theory-informed extensions, fitted to the experimental data~\cite{mackay2024countoscope}. However, in our experimental system, our dominant limitation is in resolving $\Nvar$ and not lack of long-time data (SI Fig.~\ref{fig:plateaus vs L}). 
For boxes $L \lesssim 0.3\sigma$, the correlation function decays so quickly that our imaging timestep can not capture the early time decay, and hence $T(L)$ is incorrectly estimated. To circumvent this, we match our first experimental data point $C_N(\Delta t)$ where $\Delta t = 0.5$~s, is our experimental time, with a short time-small box formula in Eq.~\eqref{eq:fluct1} obtained via the theory, and integrate the  short time extension over $[0,\Delta t]$ (see dotted coloured lines in Fig.~\ref{fig:timescale}-a and b for $t \leq 0.5$~s). This short time extension is added on all box sizes but only makes a significant difference on box sizes $L \lesssim \sigma$ where the decay of the correlation function is otherwise too fast to be captured by data.

%\sophie{where to put that? Although these two quantities are related via Eq.~\eqref{eq:nMSD}, important differences should be noted. First, because $\DNtt$ computes differences between time points and hence between images, $\DNtt$ is less sensitive to detection errors, or ``stuck'' particles -- as also suggested in \cite{mackay2024countoscope,cerbino2008differential}. Second, $\DNtt$ is more sensitive to short time evolution, while $C_N(t)$ to long time decays. Since $N(t)$ decorrelates in time with complex patterns, not single exponentials, it is useful to probe different ways of accounting for the decorrelation time through a single timescale.   }

\section{Comparison to a common Fourier-based approach}
\label{sec:comparison to fourier}

To assess the performance of the Countoscope, we compare our results to another approach to calculate $\Dcoll$ at equilibrium. Centre of mass approaches have low statistical resolution, and thus we rather focus on a Fourier-based approach for benchmarking. 

%At equilibrium, there are several established, albeit potentially difficult to implement, techniques to measure $\Dcoll$. Numerous works rely on probing the diffusion coefficient of the centre of mass of many particles $D_{\mathrm{cm}}$~\cite{obliger2016free,ariskina2022free,schlaich2024theory,ala2002collective,falk2015subcontinuum,thorneywork2015structure}. %$D_{\mathrm{cm}}$ is sometimes referred to as ``collective'' diffusion coefficient, in which case $\Dcoll$ is termed ``transport'' diffusion coefficient. 
%The collective diffusion is then obtained as $\Dcoll = D_{\mathrm{cm}}/\SkZ$. Yet, obtaining a statistically meaningful trajectory for the centre of mass requires following a large group of particles for a substantial amount of time, which is experimentally challenging as particles continuously exit and enter the field of view. Even in simulations, only one trajectory is obtained, limiting statistical resolution. Because this technique is essentially impractical and has low statistical resolution, we discard it for benchmarking, and focus on Fourier-based approaches. 

\subsection{Dynamic structure factors}

% \paragraph{Workflow of method 3.}
% \adam{don't call this method 3. call it f(k, t) or whatever}

A common approach to characterize the relaxation of diffusion processes at several scales is to analyze dynamic structure factors $F(k,t)$ for a given wavenumber $k$ after a time interval $t$. The dynamic structure factor is also referred to as the intermediate scattering function~\cite{dhont1996introduction}. Formally, it is defined in Fourier space as the correlation function of the Fourier-transformed densities, $F(k,t) = \langle \hat{\rho}(\bm{k},t) \hat{\rho}^{\star}(\bm{k},0) \rangle/N_p$ % \adam{I think this should be "the correlator of the fourier-transformed density function" or similar. The fourier transform of the density correlation function would be $\mathcal{F}\{G(r, t)\}$ but then you need the $1+\rho$ as we discussed with Tristan. Maybe having both definitions is a good idea? I can also write this if you like}. 
where $N_p$ is the number of particles of the suspension. Each particle indexed by $\mu$ has 2D coordinates given by $\bm{r}_\mu(t) = (x_\mu(t),y_\mu(t))$, and one can equivalently rewrite 
\begin{equation}
    F(k,t) = \frac{1}{N_p} \sum_{\mu, \nu = 1}^{N_p} \left\langle e^{i \bm{k} \cdot (\bm{r}_\mu(t) - \bm{r}_\nu(0))} \right\rangle
    \label{eq:fkt}
\end{equation}
where we assumed the system is rotationally invariant such that $F$ only depends on $k = |\bm{k}|$. At time zero, $F$ is equal to the static structure factor $F(k,t=0) = S(k)$. Calculating $F(k,t)$ via Eq.~\eqref{eq:fkt} is also called the direct method~\cite{sedlmeier_spatial_2011}. %\sophie{But when $N_p(t)$ varies in time because some particles are in and out of the box, how is this sum performed? only over particles that are there at both times? that doesn't sound like a great idea? How did we do it?} \adam{this is in SI section 1.3}

The dynamic structure factor $F(k,t)$ characterizes how the structure of the fluid evolves from a given state. Within linear response, \textit{i.e.} assuming density fluctuations are small, the structural dynamics are fully described by a diffusion coefficient in Fourier space $D(k,t)$ such that 
\begin{equation}
    f(k, t) \equiv \frac{F(k,t)}{S(k)} =  \exp \left( - D(k,t) k^2 t \right). 
    \label{eq:fktD}
\end{equation}
Large wavenumbers refer to motion at small scales and hence correspond to individual motion, so we expect $\Dlargekt = \Dself(t)$. The limit of small wavenumbers in turn describes collective motion, by definition, ${\Dsmallkt = \Dcoll(t)}$.

It is important to note that the dynamic structure factor bundles two contributions: correlations between a given particle at a given point in time with \textit{itself} at a later time $F_s(k,t)$, and correlations between \textit{distinct} particles at different times $F_d(k,t)$. This means $F(k,t)$ can be rewritten as 
\begin{equation}
    \begin{split}
        F(k,t) &= F_s(k,t) + F_d(k,t) \\
        &= \frac{1}{N_p} \sum_{\mu}^{N_p} \left\langle e^{i \bm{k} \cdot (\bm{r}_\mu(t) - \bm{r}_\mu(0))} \right\rangle ... \\
        & \,\,\, \,\,\,\,\,\,  +  \frac{1}{N_p} \sum_{\mu\neq\nu}^{N_p} \left\langle e^{i \bm{k} \cdot (\bm{r}_\mu(t) - \bm{r}_\nu(0))} \right\rangle.
    \end{split}
\end{equation} 
The relaxation of the self part is entirely dictated by the self diffusion coefficient, as $F_s(k,t) = \exp \left( - \Dself k^2 t \right)$. In principle, $\Dself$ depends on time. %In practice, $F_s(k,t)$ decays so quickly, at least for the large wavenumbers $k$, that one can only reliably extract the short-time self diffusion coefficient. \alice{Don't normally like to suggest self citations but, in my soft matter paper from 2016 we discuss in detail how to get both Ds and Dl out of the self ISF. So maybe that is useful here. } %\sophie{long and short time self we should be careful about again}. 

Turning now to the full dynamic structure factor, by inverting the decay of $f(k,t)$, one can obtain $D(k,t)$ linked to collective dynamics.
So far we have kept a dependence of $D(k,t)$ on time, yet to simplify data analysis, for now, we will focus on short times, i.e., $D(k) = D(k, t \simeq 0)$,  as data is generally better resolved at short times. In practice, this involves inverting Eq.~\eqref{eq:fktD} at the first (non zero) time point. %\sophie{there's a question of small vs large wavevectors here that we should be careful about}%\ref{fig:fktlongshort}).
In simulations and in the theory, we only consider hard sphere interactions, which are clearly pairwise additive and hence the collective diffusion coefficient is expected to be independent of time~\cite{pusey_liquids_1991}. In experiments, the collective diffusion coefficient also appears to be independent of time, which may be a result of density fluctuations being small at the investigated packing fractions~\cite{dhont1996introduction}. In the SI, we distinguish short and long time regimes, and show similar results overall (SI Fig.~S6).%\adam{could comment more here for reviewer 3} \adam{this might need updating now you said $\Dcoll(t=0) = \Dcoll(t=\infty)$ in the introduction}

%\vspace{3mm}

\subsection{Divergence artefact of the dynamic structure factor for $\smallk$.} \label{sec:fkt computation}

\begin{figure}[h!]
    \centering
    \includegraphics[width=0.99\linewidth]{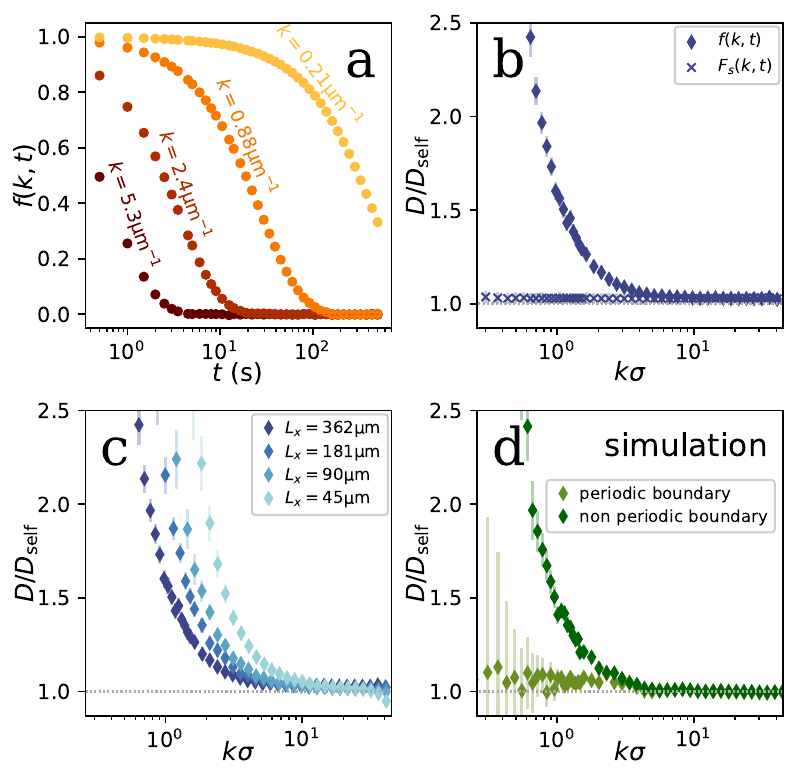}
    \caption{\textbf{Challenges in resolving Fourier space relaxation from experiments at $\phi = 0.02$.} (a) Examples of dynamic structure factors $f(k,t)$ at different wavenumbers $k$. %\sophie{whatever we do here in terms of fitting needs to reflect what we decide to keep in the paper: all times or short times only}. \adam{make sure the full x axis on (a) is shown in order to show it's fullyish deccorrelated.}  
    (b) Obtained $D(k)$ from first-point inversion of $f(k,t)$ and $F_s(k,t)$. %\adam{should we comment on disgrepancy between $F_s$ and MSD?}
    (c) $D(k)$ from $f(k,t)$, but for cropped microscopy images. $L_x$ as given in legend, $L_y$ was picked to preserve aspect ratio of the field of view. (d) $D(k)$ from $f(k,t)$ from simulations with periodic and non-periodic boundary conditions. The numerical field of views are taken to be the same size. This subplot is repeated for the dense case in SI Fig.~\ref{fig:periodicness high density}.
    Error bars are propagated from the standard error in the value of $f(k,t)$ across all time origins.
    }
    \label{fig:fourierIssues}
\end{figure}

%To make progress with the Fourier space method, we work on an example, the dilute suspension at $\phi = 0.02$. 
The dynamic structure factor $F(k,t)$ is computed via Eq.~\eqref{eq:fkt} at various wavelengths $k$  for the dilute suspension, $\phi = 0.02$, and shown in \mbox{Fig.~\ref{fig:fourierIssues}-a}. 
%To extract the scale-dependent diffusion coefficient $D(k)$, we invert . 
From this we obtain $D(k)$ for all relevant wavelengths as presented in Fig.~\ref{fig:fourierIssues}-b (diamonds). Surprisingly, for this dilute suspension we notice a clear divergence at small wavelengths of $D(k)$. Note that this divergence is not visible in the self diffusion coefficient $\Dself$ extracted in a similar way from $F_s(k,t)$ (\mbox{Fig.~\ref{fig:fourierIssues}-b}, crosses). We do not expect such changes in the collective diffusion coefficient $\Dsmallk$ for such a dilute suspension.

To unravel the origin of this peculiar artefact, we conduct the analysis again on modified versions of the experimental data. First, we trim the duration of our experimental data, and plot $D(k)$ for different trimmed lengths, and find no significant difference (see SI Fig.~S7). %\ref{fig:f divergence movie length}). 
The artefact is thus not due to a lack of statistics. Second, we crop experimental movies, effectively reducing the effective field of view $(L_x, L_y)$. The divergence in $D(k)$ is significantly affected by cropping: occurring at larger wavenumbers for smaller images, see Fig.~\ref{fig:fourierIssues}-c. Typically, the divergence starts for $k$ values such that $k \lesssim 2\pi/(L_x/10)$. This hints that the artefact originates from edge effects. Finally, we also compute $D(k)$ from simulation data, see Fig.~\ref{fig:fourierIssues}-d (light green), and do not find the divergence. 

The fundamental difference between experiments and simulations is that the simulated data has periodic boundary conditions. To mimic the experimental situation, we run a simulation on a much larger simulation box and do the $D(k)$ analysis on a cropped subset of the data, see Fig.~\ref{fig:fourierIssues}-d (dark green). Under these simulated non-periodic boundary conditions, we recover a similar divergence in $D(k)$ as found experimentally. This demonstrates that the absence of periodic boundary conditions is at the origin of the artefact. The resolution of the dynamic structure factor is thus intrinsically limited by the finite field of view of the experiment.

While a detailed investigation of why this artefact appears is beyond the scope of this work, edge effects are its roots. Edge effects are common on Fourier transforms of images, where numerous techniques have been developed to limit related artefacts~\cite{moisan2011periodic,zuccolotto2024improving}. Such techniques are not directly applicable to discrete particle positions, and hence to the computation of $F(k,t)$ via Eq.~\eqref{eq:fkt}, yet motivate perspectives for improvement. To conclude, we simply raise caution and stress that for experimental data with observation windows of size $L_x$, $k$ should be restricted such that $2\pi/k \lesssim 0.1 L_x$ when probing dynamic behaviour. %\adam{\cite{lin2014divergence} seems to go over this limit, but they don't report low density data, and for high density it diverges because of hydro anyway so the non-periodic divergence might be obscured}

\subsection{Results for $D(k)$ in a dense suspension.} 
\begin{figure}[h!]
    \centering
    \includegraphics[width=0.99\linewidth]{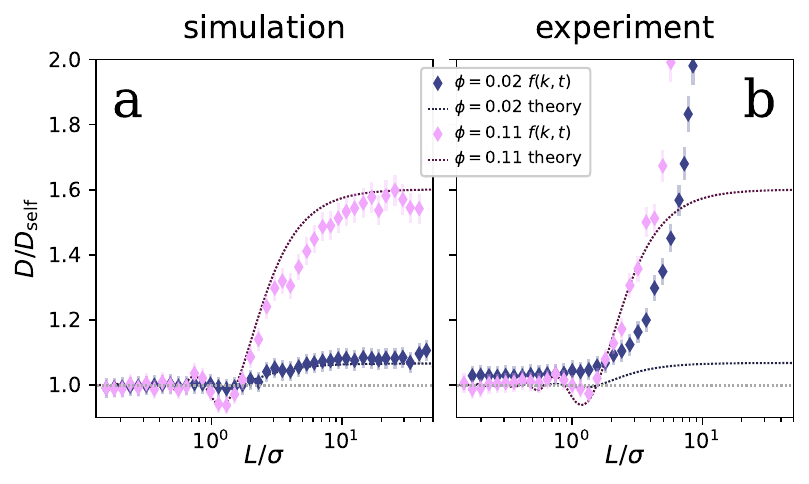}
    \caption{\textbf{Relaxation of dynamic structure factors.}
    $D(L = 2\pi/k)$ for $\phi = 0.02$ (blue) and $\phi = 0.11$ (pink) obtained from short time analysis of the dynamic structure factor in (a) simulations and (b) experiments. Theory lines correspond with  $D(k) = \Dself/S(k)$ with $S(k)$ given in Eq.~\eqref{eq:Sk}. The divergence for $L \gtrsim 3 \sigma$ in (b) corresponds to non-periodic boundaries as discussed in Fig.~\ref{fig:fourierIssues} and in the text. 
    Error bars are propagated from the standard error in the value of $f(k,t)$ across all time origins.
    }
    \label{fig:method3}
\end{figure}

We now investigate the wave-number dependent $D(k)$ for the dense suspension $\phi = 0.11$. To relate to our previous results obtained via the Countoscope, we present data as $D(k = 2\pi/L)$, effectively flipping the x axis horizontally.  Fig.~\ref{fig:method3}-a shows the results from simulation compared to the theoretical prediction $D(k) = \Dself/S(k)$ (dotted lines in Fig.~\ref{fig:method3}).
Simulations without hydrodynamics perfectly reproduce the theory, and plateau to the collective diffusion coefficient $\Dcoll$ as expected. Crucially this confirms the validity of our analysis scheme. Distinct features appear in $D(k)$ at the higher packing fraction at lower wavenumbers. In particular, we notice a minimum in $D(k)$ around $L \simeq \sigma$ ($\simeq 2\pi/k$). The scale-dependent diffusion coefficient $D(k)$ is quite sensitive to the fluid's structure at different wavelengths, as expected from $D(k) \sim 1/S(k)$. The maximum in $S(k)$ (see Fig.~\ref{fig:Sk appendix}), corresponding to ordering at increased packing fractions, thus corresponds with a minimum in $D(k)$, the so-called De Gennes narrowing~\cite{kellouai2024gennes}.  This interpretation can be checked by overlapping the theoretical prediction $D(k) = \Dself/S(k)$ on simulation and experiments (dotted lines in Fig.~\ref{fig:method3}). %, where $S(k)$ is given by Eq.~\eqref{eq:Sk}. 

In contrast, sharp discrepancies arise between experimental data and theory for large lengthscales ${L \gtrsim 3 \sigma}$ (\mbox{Fig.~\ref{fig:method3}-b}). This divergence occurs at a similar lengthscale for experimental data at both ${\phi= 0.11}$ and ${\phi= 0.02}$, suggesting it is in both cases a consequence of the artefact discussed above. As such, beyond this value of $k$, interpretation of the data is ambiguous. Nonetheless, at lower wavenumber, the distinct features linked to De Gennes narrowing for the higher packing fraction are accurately resolved in $D(k)$ for the experiment.  
Divergence of $D(k)$ in previous works has been attributed to hydrodynamic corrections~\cite{lin1995experimental,lin2014divergence,bleibel2014hydrodynamic}. Unfortunately, this reported behaviour arises for wavelengths where the divergence artefact in $D(k)$ kicks in for our results, and before the theory even reaches the $\Dcoll$ plateau. To deconvolve hydrodynamic contributions from this artefact is beyond the scope of our work. Yet, several strategies are worth mentioning. On the experimental side, one could use wider fields of view to increase the range of relevant $L$ lengths or design de-aliasing techniques~\cite{zuccolotto2024improving,moisan2011periodic}. Simulations including hydrodynamic interactions between particles could be conducted~\cite{sprinkle2020driven,mackay2024countoscope} and compared with simulations with purely steric interactions as well, but require intense computational resources. %: only long simulation timescales and wide simulation boxes can resolve decorrelation accurately at these large wavelengths. 

%we compare our data to simulations with and without hydrodynamics -- Fig.~\ref{fig:method3}-b. Because hydrodynamic simulations reproduce experimental data so well at all scales in the counting framework \sophie{check and reference figure}, they serve as a faithful proxy to assess the role of hydrodynamics. 
%curves as $D(k) = \Dself/S(k)$ confirming that we operate with numerical parameters -- numerical box size, total simulation time -- that are sufficient to resolve this discrepancy. Simulations with hydrodynamics do not visibly diverge over the range $L \simeq 5 - 20 \sigma$, unlike experimental data. For larger wavelengths, the data is so noisy that it is not possible to assess the presence of the divergence. One can therefore conclude that at least over a finite range of wavelengths, the divergence in experiments is an artefact. Hydrodynamics, however, do play a role at several wavelengths. In particular at intermediate wavelengths, $L \simeq 2-10 \sigma$, they appear to slow down diffusion, while at larger wavelengths, they speed them up. 

\subsection{Countoscope versus Fourier-based approaches}

\begin{figure}[h!]
    \centering
    \includegraphics[width=0.99\linewidth]{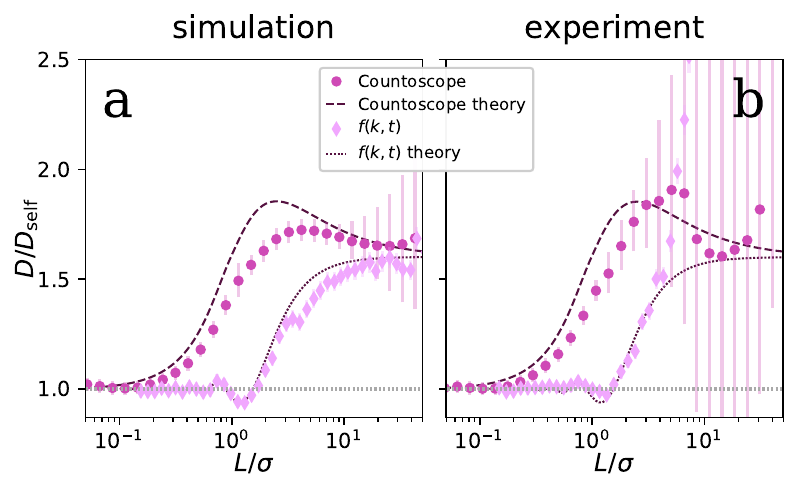}
    \caption{\textbf{Comparison of Countoscope versus Fourier-space approaches:} via $D(L)$ computed via the timescale integral (dots) and $D(k = 2\pi/L)$ computed via the early-time fit of the dynamic structure factor (diamonds), for (a) simulations and (b) experiments for the highest packing fraction $\phi = 0.11$. Theory lines correspond with Eq.~\eqref{eq:Dbox} for the Countoscope and $D(k) = \Dself/S(k)$ with $S(k)$ given in Eq.~\eqref{eq:Sk} for the Fourier approach. %\sophie{Also waiting for updated data where sim data should converge more nicely to the plateau, and exp. countoscope data should follow theory more closely.}
    }
    \label{fig:FinalComp}
\end{figure}

The features observed in $D(k)$ via the investigation of the dynamic structure factor are reminiscent of features of $D(L)$ inferred from the Countoscope. We compare the two approaches by overlapping $D(k)$ and $D(L)$ for simulations and experiments in the dense regime $\phi = 0.11$ (Fig.~\ref{fig:FinalComp}). 
Overall, the fluid's structure is quite apparent on $D(k)$, and less so on $D(L)$. Curiously the maximum in $D(L)$ occurs at a similar lengthscale to the minimum in $D(k)$, a behaviour which would require further investigation at different packing fractions to be confirmed. 
The increase in the scale-dependent diffusion coefficient occurs at different scales: counting is sensitive to collective effects typically for $L \gtrsim \sigma$ while one must wait for $2\pi/k \gtrsim 5\sigma$ for collective effects in $D(k)$, demonstrating the sensitivity of counting. 

Based on simulation data, we find counting can estimate the collective diffusion coefficient $\Dcoll$ on large boxes. Indeed, in Fig.~\ref{fig:FinalComp}-a, we observe $\Dsmallk = \DlargeL = \Dcoll$. The measurement of the collective diffusion coefficient $\Dcoll$ via the Countoscope or the dynamic structure factor approach are thus equivalent. As a periodic methodology, the Fourier-based approach on simulation data does not appear to diverge at large length-scales, compared to the Countoscope. Although the divergence is not apparent at the length scales presented in Fig.~\ref{fig:FinalComp}-a, it is apparent at larger length scales (SI Fig.~\ref{fig:D periodic size}). Nonetheless, one should keep in mind that finite-simulation size effects can still affect the plateau reached by $D(k)$ significantly (SI Fig. \ref{fig:D periodic size}). As a Fourier methodology, dynamic structure factors access larger lengthscales than counting on periodic simulation data. 

In experiments, it is hard at this stage to compare the limiting behaviour of $D(L)$ and $D(k)$ for $\largeL$ or $\smallk$ due to the artefact divergence in $D(k)$. However, in experiments, counting provides information on $D(L)$ at much larger lengthscales than $D(k)$. Improving variance estimates would eventually increase the timescale integral's accuracy and decrease errors from the counting technique at the largest lengthscales. For $D(k)$, however, the system will always stay non-periodic and the divergence may be harder to fix. As a real-space methodology, counting is more robust on non-periodic experimental data than Fourier-based approaches. 

\section{Discussion and Conclusion}
\label{sec:discussion and conclusion}

%\sophie{add back in something in the SI}

%In addition to direct interactions set by the interparticle potential, the suspending fluid also introduces hydrodynamic interactions - long range particle-particle interactions mediated by the fluid. %\sophie{text}
%Particle interactions are generally more complex than simply repulsive and include at least interactions mediated by the suspending fluid, or hydrodynamic interactions. 
%However, even for suspensions of hard sphere-like colloids, experimental results widely differ in assessing the role of hydrodynamic interactions on collective properties~\cite{panzuela2018solvent, panzuela2017collective,kops1982dynamic,qiu1990hydrodynamic, segre1995experimental, bleibel20153d, lin1995experimental,lin2014divergence,bleibel2014hydrodynamic,bleibel20153d, falck2004influence,mackay2024countoscope}. Such discrepancies arise from the challenge of m

%\sophie{---}

In this work, we have shown how to infer collective diffusion properties at various spatial lengthscales by counting particles in boxes. The box-size dependent diffusion coefficient $D(L)$ can be obtained via time integration of the correlation function of particle numbers in a box $\langle N(t) N(0) \rangle$. $D(L)$ converges to the self diffusion coefficient in small boxes $D(L\rightarrow 0) = D_{\rm self}$. In large boxes, it probes the collective diffusion coefficient ${D(L \rightarrow\infty) = \Dcoll}$, as confirmed via a comparison between theory and simulations and investigation of the dynamic structure factor. 
For experiments, accurately determining $D(L)$ for large length scales is hindered by difficulties in determining the variance of the number fluctuations. However, we have found Fourier approaches, such as the dynamic structure factor, also struggle at large length scales due to the finite field of view of microscopy images, which leads to unphysical divergences of $D(L)$ at large wavelengths. In contrast, counting exploits finite fields of view by deliberately paving the image with finite observation boxes and could be made more accurate with an improved estimate of $\Nvar$. Finally, $D(L)$ informs on collective dynamics at all spatial scales $L$, allowing us to broadly investigate collective properties of suspensions. 

While our investigation was centred on a 2D colloidal suspension, it introduces a general tool to infer the collective properties of suspensions from microscopy images. Indeed, the formalism applies to any quasi-2D experimental scenarios where particle coordinates can be obtained. Beyond that, we anticipate the Countoscope to be applicable in 3D. Given 3D particle coordinates, one can readily calculate number fluctuations in a 3D box, and then perform correlations, and the timescale integral. Interpretation of the results would be facilitated with further theory, which is \textit{a priori} feasible with the current theoretical model~\cite{minh2023ionic}. However, 3D experimental setups often involve further complexities which remain to be accounted for in this context, and we leave this for further investigation.

\begin{table*}
\centering
\begin{tabular}{c|cccc}
%\hline
 & \multicolumn{1}{c}{scattering} & \multicolumn{3}{c}{microscopy}
 \\
 & $\overbrace{\hspace{0.2cm} \text{intensities} \hspace{0.2cm}}$& \multicolumn{3}{c}{$\overbrace{\hspace{0.4cm} \text{intensities \hspace{1.0cm} positions \hspace{0.8cm} trajectories} \hspace{0.2cm} }$}
 \\
%\cline{2-5}
 & \includegraphics[width=0.1\textwidth]{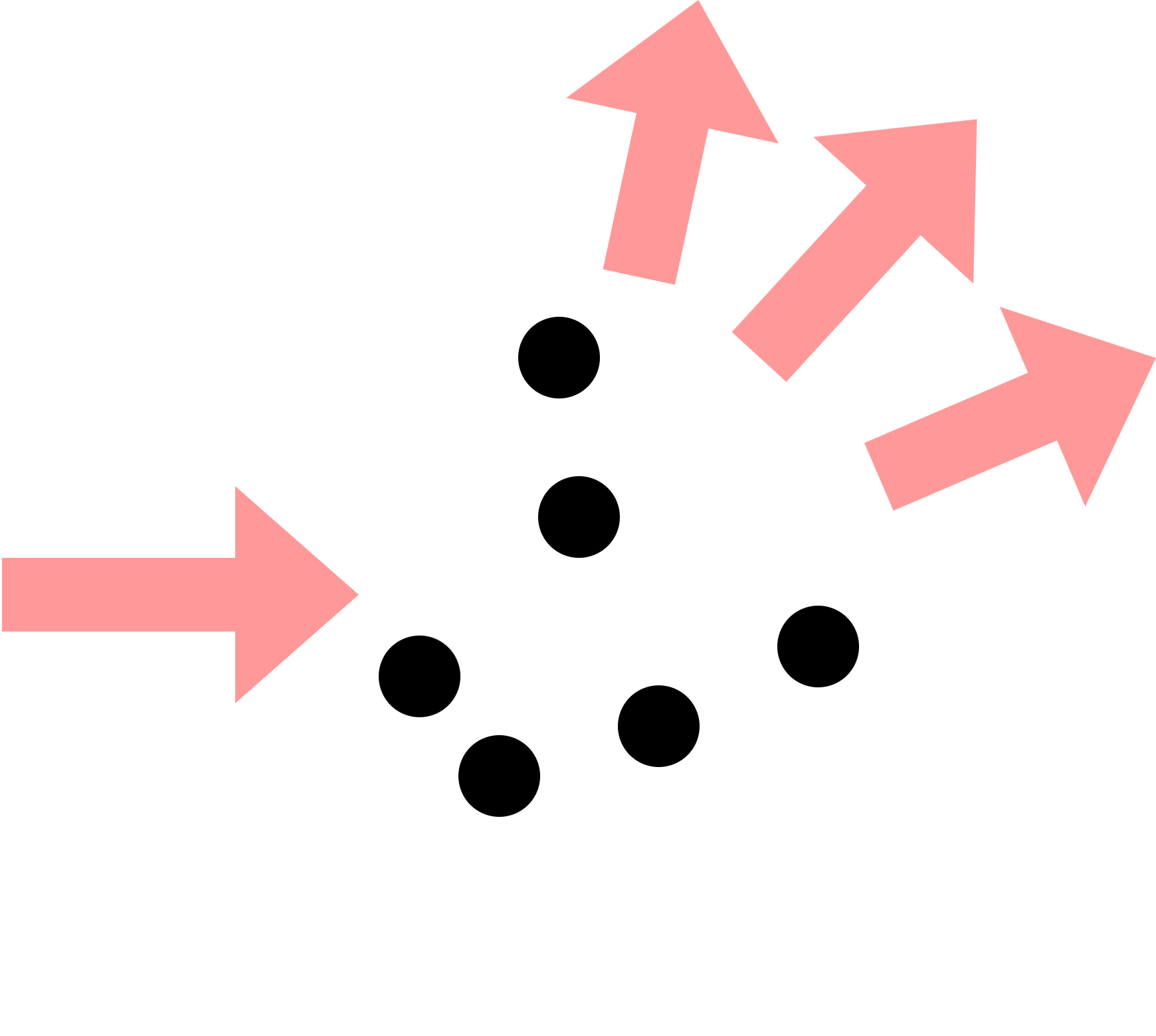} & \includegraphics[width=0.1\textwidth]{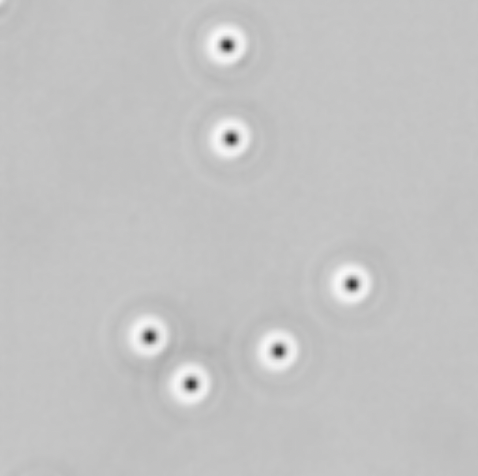} & \includegraphics[width=0.1\textwidth]{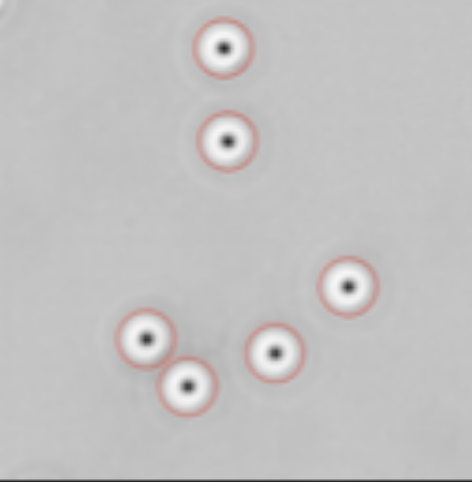} & \includegraphics[width=0.1\textwidth]{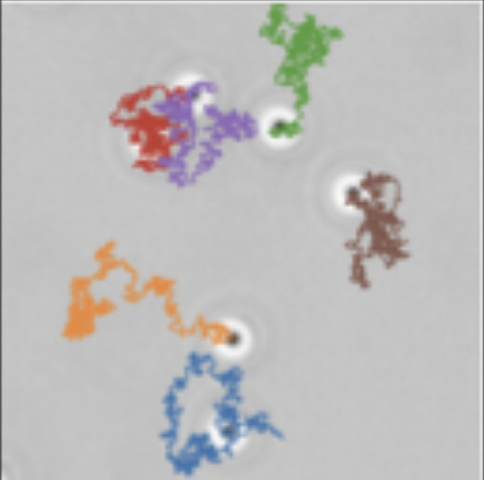} \\
\hline
Real-space & FCS, DLS & \textcolor{gray}{Intensity Counting} & Countoscope & MSD \\
Fourier-space &  XPCS, DLS & DDM & $F(k, t)$ & $F_s(k, t), F_d(k, t)$ \\
%\hline
\end{tabular}
\caption{Comparison of different techniques for analyzing particle dynamics in scattering and microscopy experiments. FCS = Fluorescence Correlation Spectroscopy \cite{elson1974fluorescence}, DLS = Dynamic Light Scattering \cite{berne2000dynamic}, XPCS = X-ray Photon Correlation Spectroscopy, DDM = Dynamic Differential Microscopy \cite{cerbino2008differential}, MSD = Mean Squared Displacement}
\label{tab:techniques}
\end{table*}

We anticipate that our method could shed light on the effect of hydrodynamic interactions in colloidal suspensions, in particular for quasi 2D suspensions near walls, which occur quite commonly in soft matter systems. Previous theoretical and experimental investigations of quasi 2D geometries have suggested that long-range correlations between particles can enhance collective motion dramatically, resulting in a divergence of $D(k)$ at large wavelengths~\cite{lin1995experimental,lin2014divergence,bleibel2014hydrodynamic,bleibel20153d,pelaez2018hydrodynamic,falck2004influence,nagele_dynamic_2002,pesche2000stokesian}. This is at odds with bulk 3D systems, where hydrodynamic interactions reduce the value of $\Dcoll$~\cite{qiu1990hydrodynamic, segre1995experimental, bleibel20153d}. 
Interestingly, we find in our experimental data that a divergence on $D(k)$ can be linked to edge effects, and thus potential effects of hydrodynamic interactions are hidden. In contrast, on experimental data for $D(L)$, we find no evidence for a divergence, however the error bars in our data at large length scales are very large. %we appear to reach a plateau in $D(L)$ with our Countoscope approach, at least over a certain range of length scales 
%\adam{is this meant to say simulation? We don't really reach a plateau in $D(L)$ on experiments}\alice{no but we also don't see a divergence - suggested alternative}. 
%Our experimental data suggests a limiting $\Dcoll$ which is only slightly higher than the one described by simulations or theory without hydrodynamic interactions \adam{I find this rather far fetched, our experimental data is too variable to reliably compare to the theory.}.
%\adam{complicating this, I now have hydro simulation data that makes a very strong case that there is a divergence}
Beyond the artefact divergence, which could also be at play in previous works, discrepancies could originate from diverse physical factors. At higher packing fractions, up to $\phi \simeq 0.6$, we expect hydrodynamic effects could be more important~\cite{mackay2024countoscope}. The geometry of our system consists of particles at a single wall as opposed to a fluid-fluid interface or between two closely-spaced walls~\cite{lin1995experimental,lin2014divergence,bleibel2014hydrodynamic,bleibel20153d,pelaez2018hydrodynamic,falck2004influence}, and geometry is known to significantly influence the range over which hydrodynamic interactions decay~\cite{hashemi2022computing,Liron1976}.
The decay of hydrodynamic interactions could also be a transient short-time effect, requiring further interpretation of $\Dcoll$ as a time-dependent property~\cite{panzuela2018solvent}. In more complex cases such as these, it remains an open question to identify whether fluctuating counts are sensitive to short or long time collective properties.

More generally, we stress that to obtain accurate quantification of collective properties, significant data sets are required, both in time and spatially: our experimental data sets are 100 times larger than the particle size, and more than 1000 times longer than the time to diffuse across a particle's diameter and only nearly capture $\Dcoll$.
%The counting recipe provides further tools to investigate collective phenomena, and understand the effects of hydrodynamic interactions in various geometries. 
Given that wide fields of view are necessary to resolve motion at large spatial scales, % Especially in dense systems where collective effects arise, this means that 
for dense systems in particular trajectory reconstruction may no longer be feasible~\cite{crocker1996methods,manzo2015review,deforet2012automated}. For the investigation of collective effects, this is not an issue as both counting or dynamic structure factors $F(k,t)$ do not require trajectories (Table~\ref{tab:techniques}). In that sense, counting fills a gap in the field, as the real-space equivalent of $F(k,t)$. Density fluctuations are also investigated in real space through intensity fluctuations of scattered light, via Fluorescence Correlation Spectroscopy (FCS)~\cite{elson1974fluorescence} or Dynamic Light Scattering (DLS)~\cite{berne2000dynamic}. In contrast with these techniques, we explicitly count numbers, avoiding the link between scattered intensities and particle numbers which is especially ambiguous at high densities~\cite{hassan2015making,rose2020particle}. More importantly, the Countoscope is not restricted to a given lengthscale, unlike \textit{e.g.} FCS which analyzes the scattered light of a given illuminated region. %Here, the different scales are explored in the post-analysis and not constrained during the experiment. 
This suggests exploring intensity correlations in real space on virtual boxes of an image, a form of ``intensity Countoscope''. Again, this would fill a gap in the field, as the real space equivalent of Differential Dynamic Microscopy~\cite{cerbino2008differential}. 

%\sophie{Somewhere ... Generally, we stress that to obtain accurate quantification of collective properties, significant data sets are required, both in time and spatially, because collective modes can extend over large time and length scales. Therefore, the dominant limitation to resolve collective properties may not lie in the use of a specific method, but rather on acquiring sufficiently large datasets. }

Finally, we hypothesize that probing number correlations at different scales could inform us about more diverse collective transport properties, beyond diffusion. For instance, in active matter systems, either synthetic or biological, peculiar features are common in static number fluctuations: ``Giant'' number fluctuations, where $\alpha > 0$ in the scaling $\langle N ^2 \rangle - \langle N \rangle^2 \sim N^{1 + \alpha}$, indicate long-range organization, as found in bacterial aggregates and active matter~\cite{zhang2010collective,peruani2012collective,liu2021density,fily2012athermal,alarcon2017morphology,chate2008collective,fadda2023interplay,narayan2007long,toner2005hydrodynamics}. 
Likely, investigating the dynamic counterpart of these static fluctuations, through the decorrelation time $T(L)$ of number fluctuations at different scales could help us characterize collective motile states, and perhaps shed light on how they emerge from specific interparticle interactions~\cite{liebchen2021interactions,dijkstra2021predictive}. 

\section*{Data and code availability}

All data needed to evaluate the conclusions in the paper are present in the paper and/or the Supplementary Materials. All other data are available upon reasonable request to the authors.

Code for the Countoscope is available \cite{sophie_marbach_2025_15000583}. Code for computing the dynamic structure factor is also available \cite{adam_2025_14999376}.

%\section*{Code availability}

%Upon publication the code to analyze particle number fluctuations will be made available on Github. Simulation codes to generate particle trajectories are available at \url{https://github.com/stochasticHydroTools/RigidMultiblobsWall/tree/master/Lubrication}

\section*{Acknowledgements}

We wish to acknowledge fruitful discussions with Jean-Louis Barrat, Roxanne Berthin, Ludovic Berthier, Roberto Cerbino, Benoit Coasne, Rafael Delgado Buscalioni, Aleksandar Donev, Simon Gravelle, Pierre Levitz, Grace Mattingly, Ama\"{e}l Obliger and Alexander Schlaich.
%Ludovic Berthier, Adam Carter, Aleksandar Donev, Roel Dullens, Daan Frenkel, Simon Gravelle, Jean-Pierre Hansen, Pierre Illien, Marie Jardat, Pierre Levitz, Th\'{e} Hoang Ngoc Minh, Ignacio Pagonabarraga and Benjamin Rotenberg. S.M. is especially grateful to Federico Paratore for a very helpful conversation on an exploratory cruise in California. B.S. and S.M. thank Johanna McCombs for her early exploratory work on this subject and Sarah A. Hughes for her invaluable help with figures. 

Financial support for this project was provided by the Institute of Materials Science (iMAT) of the Alliance Sorbonne Université.
A.C. acknowledges iMAT for a PhD grant.
A.L.T. acknowledges funding from a Royal Society University Research Fellowship (URF\textbackslash R1\textbackslash211033). E.K.R.M. and A.L.T acknowledge funding from EPSRC (EP/X02492X/1).  B.S. acknowledges funding from the National Science Foundation award DMS-2052515.  %\sophie{Please fill in}

\section*{Author contributions}

The authors confirm their contribution to the paper as follows: 
study conception and design: S.M.; simulation data collection: A.C.; simulation design: A.C., B.S.; experimental data collection: E.K.R.M.; experimental design: E.K.R.M., A.L.T., modeling: A.C., S.M.; data analysis: A.C.; data interpretation: A. C., B.S., A.L.T., S.M.; visualization: A.C.; draft manuscript preparation: A.C., S.M.; review and editing: A.C., A.L.T, S.M.

\section*{Competing interests}

The authors declare no competing interests.

%\bibliography{collective}% Produces the bibliography via BibTeX.

%\adam{make sure the arxiv version has hyperlinks in the references!} \sophie{You can be my guest to have all the fun in the world adding hyperrefs to all of these! we can post the arxiv a couple days after we send it out to soft matter. }
%\adam{haha sure, it's a personal bugbear of mine when references are not linked in 2024. What do you make your .bib with? I think zotero will do it automatically}


%\newpage

%\nocite{*}

\end{document}